\documentclass[aps,prx,reprint,showpacs,twocolumn,superscriptaddress,eqsecnum]{revtex4-2}
\usepackage{xr}
\usepackage{amsmath}
\usepackage{amsfonts,amssymb,amsthm,amsxtra}
\usepackage{algorithm} 
\usepackage{algpseudocode} 
\usepackage[]{graphicx}
\usepackage{mathrsfs}
\usepackage{bbm}
\usepackage{bm}
\usepackage{braket}
\usepackage{xcolor}
\usepackage{mathtools}
\usepackage{tikz-cd}
\usepackage{lipsum}
\usepackage{booktabs}
\usepackage{mhchem}
\usepackage{float}
\usepackage[bookmarks=false,colorlinks=true,urlcolor=blue,citecolor=blue,linkcolor=blue]{hyperref}

\newtheorem*{theorem*}{Theorem}

\begin{document}

\title{Experiments with the 4D Surface Code on a QCCD Quantum Computer}

\author{Noah Berthusen}
\email{nfbert@umd.edu}
\affiliation{Joint Center for Quantum Information and Computer Science, NIST/University of Maryland, College Park, Maryland 20742, USA}
\affiliation{Quantinuum, Broomfield, CO 80021, USA}
\author{Joan Dreiling}
\author{Cameron Foltz}
\author{John P. Gaebler}
\author{Thomas M. Gatterman}
\author{Dan Gresh}
\author{Nathan Hewitt}
\author{Michael Mills}
\author{Steven A. Moses}
\author{Brian Neyenhuis}
\author{Peter Siegfried}
\author{David Hayes}
\email{david.hayes@quantinuum.com}
\affiliation{Quantinuum, Broomfield, CO 80021, USA}

\begin{abstract}
Single-shot quantum error correction has the potential to speed up quantum computations by removing the need for multiple rounds of syndrome extraction in order to be fault-tolerant.
Using Quantinuum's H2 trapped-ion quantum computer, we implement the $[[33,1,4]]$ 4D surface code and perform the first experimental demonstration of single-shot quantum error correction with bare ancilla qubits.
We conduct memory experiments comparing the 2D and 4D surface codes and find that despite differences in qubit use and syndrome extraction circuit depth, the 4D surface code matches or outperforms the 2D surface code in both the fault-tolerant and single-shot regimes.
\end{abstract}

% \date{}

\maketitle

% \tableofcontents

\section{Introduction}

To facilitate the execution of large-scale fault-tolerant circuits, quantum error correction will be necessary. After redundantly encoding the quantum information in a quantum error correcting code (QECC), errors on the physical qubits can be detected and corrected by measuring the stabilizers of the codespace. In the ideal case, these measurements are perfect, and the resulting syndrome information is reliable. In reality, however, measuring the syndromes is itself a noisy process and can result in syndromes which do not accurately represent the errors acting on the system. The standard solution to this problem is to perform repeated stabilizer measurements and provide several rounds of syndrome information to the decoder.
Typically, a number of measurement repetitions scaling like $\Theta(d)$, where $d$ is the distance of the code, is required to be fault-tolerant to both qubit and measurement errors. Here a \textit{measurement} refers to a syndrome extraction circuit, as opposed to a measurement at the physical layer.

It was shown in Ref.~\cite{Bomb_n_2015} that for certain codes, only a single round of noisy stabilizer measurements is necessary to display increased error suppression with larger blocklengths. As such, using these \textit{single-shot} codes can significantly reduce the time overhead for quantum error correction. A number of code families have been shown to support single-shot error correction include: 3D gauge color codes~\cite{Bomb_n_2015, Brown_2016}, 4D surface codes~\cite{Dennis_2002, Duivenvoorden_2019, Higgott_2023}, quantum expander codes~\cite{Leverrier_2015}, and quantum Tanner codes~\cite{Gu_2024}, among others.
This property can be facilitated by a confinement of the residual error after decoding~\cite{Quintavalle_2021} or by linear dependencies in the stabilizer checks~\cite{Campbell_2019}.

In this work, we implement the $[[33,1,4]]$ 4D surface code on Quantinuum's H2 trapped-ion computer~\cite{moses2023}. Trapped-ion quantum charge-coupled device (QCCD) processors~\cite{Wineland98, Pino2021}, such as Quantinuum's H-series devices, have effective all-to-all connectivity facilitated by ion-transport operations. 
This allows for much easier implementation of nonlocal QECCs including the 4D surface code and other recent experimental examples~\cite{hong2024entangling, burton2024genons} than would otherwise be possible on architectures with limited connectivity.
Due to linear dependencies in its stabilizer checks, the 4D surface code has an inherent robustness to syndrome errors which make it so repeated syndrome measurements are not required for the code to display a threshold.
Notably, the code is a self-correcting quantum memory~\cite{Dennis_2002, alicki2010} and is thermally stable in a sufficiently cold environment.
To the best of our knowledge, this is the first hardware demonstration of single-shot QEC with bare ancilla syndrome extraction. Other hardware QEC experiments have performed single-shot QEC through Knill~\cite{Knill2005, ryananderson2024} and Steane~\cite{steane1997, postler2023} style syndrome extraction.
To illustrate the potential benefits of using nonlocal, single-shot codes, we perform memory experiments using 53 of the 56 qubits currently available on the H2 quantum computer.

The paper is organized as follows. In Section~\ref{sec:background} we give some background on classical/quantum error correction and its connection to $\mathbb{F}_2$-homology. We additionally introduce the 4D surface code, its properties, and single-shot QEC in general. Section~\ref{sec:experiments} describes the results of memory experiments done on Quantinuum's H2 trapped-ion quantum computer with the 4D surface code.
We conclude in Section~\ref{sec:discussion} with a brief discussion. 

\section{Background}
\label{sec:background}

\subsection{Quantum error correction}

An $[n,k,d]$ binary linear classical code $\mathcal{C}$ is a $k$-dimensional subspace of the full $n$-dimensional vector space over $\mathbb{F}_2$. Vectors in this subspace, called codewords, are elements of $\ker H$, where $H$ is the \textit{parity check matrix} (pcm).  The minimum Hamming weight of a nonzero codeword is the distance $d$ of the code.

\textit{Stabilizer codes}~\cite{Gottesman_1997} are a class of quantum error correcting codes defined to be the joint $+1$-eigenspace of an abelian group $\mathcal{S}$ called the stabilizer. For an $[[n,k,d]]$ code, the stabilizer is generated by $m=n-k$ linearly independent generators which are elements of the Pauli group on $n$ qubits, $\mathcal{P}_n$. The distance $d$ is the minimum weight of a Pauli operator that commutes with everything in $\mathcal{S}$ but is not itself in $\mathcal{S}$.
Calderbank-Shor-Steane (CSS)~\cite{Calderbank_1996, steane1996} codes are a widely used class of stabilizer codes defined by two binary linear classical codes $\mathcal{C}_X, \mathcal{C}_Z$. The pcm of $\mathcal{C}_X$, $H_X$, denotes stabilizer generators which are tensor products of $X$ and $I$. Similarly, the pcm of $\mathcal{C}_Z$, $H_Z$, defines generators that are tensor products of $Z$ and $I$. To ensure that the $X$- and $Z$-type generators commute we require that $\mathcal{C}_Z^\bot \subseteq \mathcal{C}_X$, or equivalently $H_Z^T H_X = 0$.

Measuring the eigenvalues of the stabilizer generators provides a classical syndrome which is used by a decoder to identify and prescribe corrections for potential errors in the system.
In the absence of errors, the encoded state is in the joint $+1$-eigenspace of every generator, yielding a zero syndrome. Alternatively, the presence of an error results in a nonzero syndrome bit for every generator with which it anticommutes.

\subsection{Chain complexes and $\mathbb{F}_2$-homology}

In this section, we will give a brief introduction to chain complexes, $\mathbb{F}_2$-homology, and the relationship to classical and quantum codes. We refer the reader to Refs.~\cite{Breuckmann2021, Zeng_2019} for a more complete introduction to homological algebra and its application to quantum error correction.

A \textit{chain complex} is a collection of vector spaces over $\mathbb{F}_2$ together with linear maps $\partial_i$,
\begin{equation}
    \begin{tikzcd}[every matrix/.append style={},
    every label/.append style = {font = \footnotesize},
    row sep=large, column sep=small]
        C = C_n \arrow[r, "\partial_n"] & C_{n-1} \arrow[r, "\partial_{n-1}"] & ... \arrow[r, "\partial_2"] & C_1 \arrow[r, "\partial_1"] & C_0,
    \end{tikzcd}
\end{equation}
where $\partial_{i+1} \partial_i = 0$. We refer to elements of $C_i$ as $i$-chains, elements of $Z_i(C) = \ker \partial_i$ as $i$-cycles, and elements of $B_i(C) = \text{im} \ \partial_{i+1}$ as $i$-boundaries. The $i$-th homology is then defined as the vector space of $i$-cycles modulo $i$-boundaries, 
\begin{equation}
    \label{eq:homology_group}
    H_i(C) = Z_i(C) / B_i(C).
\end{equation}
Similarly, we also define elements of $Z^i(C) = \ker \partial_{i+1}^T$ as $i$-cocycles, elements of $B^i(C) = \text{im} \ \partial_i^T$ as $i$-coboundaries, and the $i$-th cohomology, $H^i(C) = Z^i(C) / B^i(C)$.

An $[n,k,d]$ classical error correcting code can be considered a 2-term chain complex, 
where its boundary map $\partial_1 = H$ is a linear map from $\mathbb{F}_2^n$ to the vector space of syndromes, $\mathbb{F}_2^{n-k}$ (assuming $H$ is full-rank).
A CSS code can similarly be represented as a 3-term chain complex,
\begin{equation}
    \begin{tikzcd}[every matrix/.append style={},
    every label/.append style = {font = \footnotesize},
    row sep=large, column sep=small]
        C = C_2 \arrow[r, "\partial_2"] & C_1 \arrow[r, "\partial_1"] & C_0,
    \end{tikzcd}
    \label{eq:CSS_code}
\end{equation}
where $\partial_2=H_Z^T$ and $\partial_1=H_X$. The condition that $\partial_{i+1}\partial_i = 0$ translates to the requirement that the $X$ and $Z$ checks must commute, e.g. $H_Z^T H_X = 0$, and so Eq.~\eqref{eq:CSS_code} defines a valid CSS code. 
Given an arbitrary length chain complex, one may define a CSS code by only considering two consecutive boundary operators.
When identifying qubits with elements of $C_i$, the resulting code parameters are $n = \dim C_i$, $k = \dim H_i(C) = \dim H^i(C)$, and $d$ is the minimum Hamming weight of a non-trivial element in $H_i(C)$ or $H^i(C)$.
Quantum error correcting codes, i.e. 3-term chain complexes, can be obtained by taking the homological~\cite{Bravyi2014}, or hypergraph~\cite{Tillich_2014} product of two 2-term chain complexes (see Appendix~\ref{apx:code_construction}). 

\subsection{4D surface code}

The 4D surface code~\cite{Dennis_2002}, also known as the tesseract code, is, at its simplest, a higher-dimensional version of the 2D surface code. 
From this abstraction we can construct the 4D surface code by considering a hypercubic lattice and placing a qubit on each of the faces. The $Z$- and $X$-type stabilizers of the code are then associated with cubes and edges, respectively.
Alternatively, it also has a natural representation as a 5-term chain complex obtainable through repeated tensor products (see Appendix~\ref{sec:4dsurfaceconstruction}). This latter representation is the one we choose to focus on throughout this work. 
While a 4D topological code embedded into a 4D space has local stabilizer checks, embedding the code into 2D (or the 1D layout of the H2 quantum computer) requires long-range connectivity. 
Consequently, implementing the 4D surface code on hardware is currently only practical for architectures that support nonlocal gates, such as neutral atom arrays~\cite{Bluvstein_2023} and ion-traps~\cite{Wineland98, Pino2021}.  
For a hypercubic lattice with side-length $L$, we obtain a 4D surface code with the parameters $[[6L^4 - 12L^3 + 10L^2 - 4L + 1,1,L^2]]$. For $L=2$, we obtain the $[[33,1,4]]$ 4D surface code.

\subsection{Single-shot QEC}
\label{sec:single-shot}
The main draw of the 4D surface code is that it is \textit{single-shot}~\cite{Bomb_n_2015}. Normally, to be fault-tolerant (FT) to both data qubit and measurement errors, a distance $d$ code will require $\Theta(d)$ rounds of syndrome measurements. Alternatively, single-shot codes require only a single round of noisy syndrome measurement. For architectures that can support it, single-shot QEC offers the potential to improve the logical clock speed by $\Theta(d)$, which can be significant for large scale problems such as factoring and quantum chemistry where theoretical studies indicate that $d>10$ or $d>30$ depending on the underlying assumptions about the physical hardware~\cite{Beverland:2022rpv}.

In the 4D surface code, single-shot QEC is facilitated by dependencies in the stabilizer generator which allow for correction of the syndrome before decoding. In the language of chain complexes, the 4D surface code can be represented as a 5-term complex as shown below,
\begin{equation}
    \begin{tikzcd}[every matrix/.append style={},
    every label/.append style = {font = \footnotesize},
    row sep=large, column sep=small]
        C = C_4 \arrow[r, "\partial_4"] & C_3 \arrow[r, "\partial_3"] & C_2 \arrow[r, "\partial_2"] & C_1 \arrow[r, "\partial_1"] & C_0.
    \end{tikzcd}
    \label{eq:4dchain}
\end{equation}
By associating qubits with elements of $C_2$, we obtain a 
CSS code with pcms $\partial_3 = H_Z^T$ and $\partial_2 = H_X$. Additionally, we identify $\partial_4 = M_Z^T$ and $\partial_1 = M_X$ as the pcm of a binary linear code for the syndromes of $Z$- and $X$-type generators, respectively. We call these pcms \textit{metachecks} of the CSS code. Due to the constraints of the boundary maps, $\partial_{i+1}\partial_i = M_ZH_Z = M_XH_X = 0$. Hence valid $Z~(X)$ syndromes are in $\ker M_Z~(\ker M_X)$. In the presence of a measurement error affecting a syndrome, we will get a nonzero \textit{metasyndrome} which we can use to correct the syndrome before attempting to decode. This technique of ``repair syndrome" decoding was used to show competitive thresholds for the 4D surface code~\cite{Duivenvoorden_2019, Higgott_2023}. Any code which can be represented as a 5-term chain complex similarly is single-shot through metachecks~\cite{Campbell_2019}. Additionally, codes that can be described with a 4-term chain complex, such as the 3D surface code~\cite{Vasmer_2019}, are single-shot for a single type of errors.

\subsection{Decoding}
\label{sec:decoding}

To perform circuit-level decoding in this work, we use Stim~\cite{Gidney_2021} to generate a \textit{detector error model} (DEM) for a given circuit and the quantum hardware noise model. A DEM is a bipartite graph with check nodes corresponding to \textit{detectors}---parities of time adjacent syndromes---and bit nodes corresponding to possible errors in the circuit. A detector and error are connected by an edge if the error flips the corresponding detector.
This bipartite graph is used as input along with a syndrome for a decoder such as BP+OSD~\cite{Panteleev_2021, Roffe_2020, Roffe_LDPC_Python_tools_2022}, which then outputs a set of errors with their locations specified in spacetime. Decoding is considered a success if applying the guessed error has the same logical outcome as the measured observable.

\begin{figure}
    \centering
    \includegraphics[width=\linewidth]{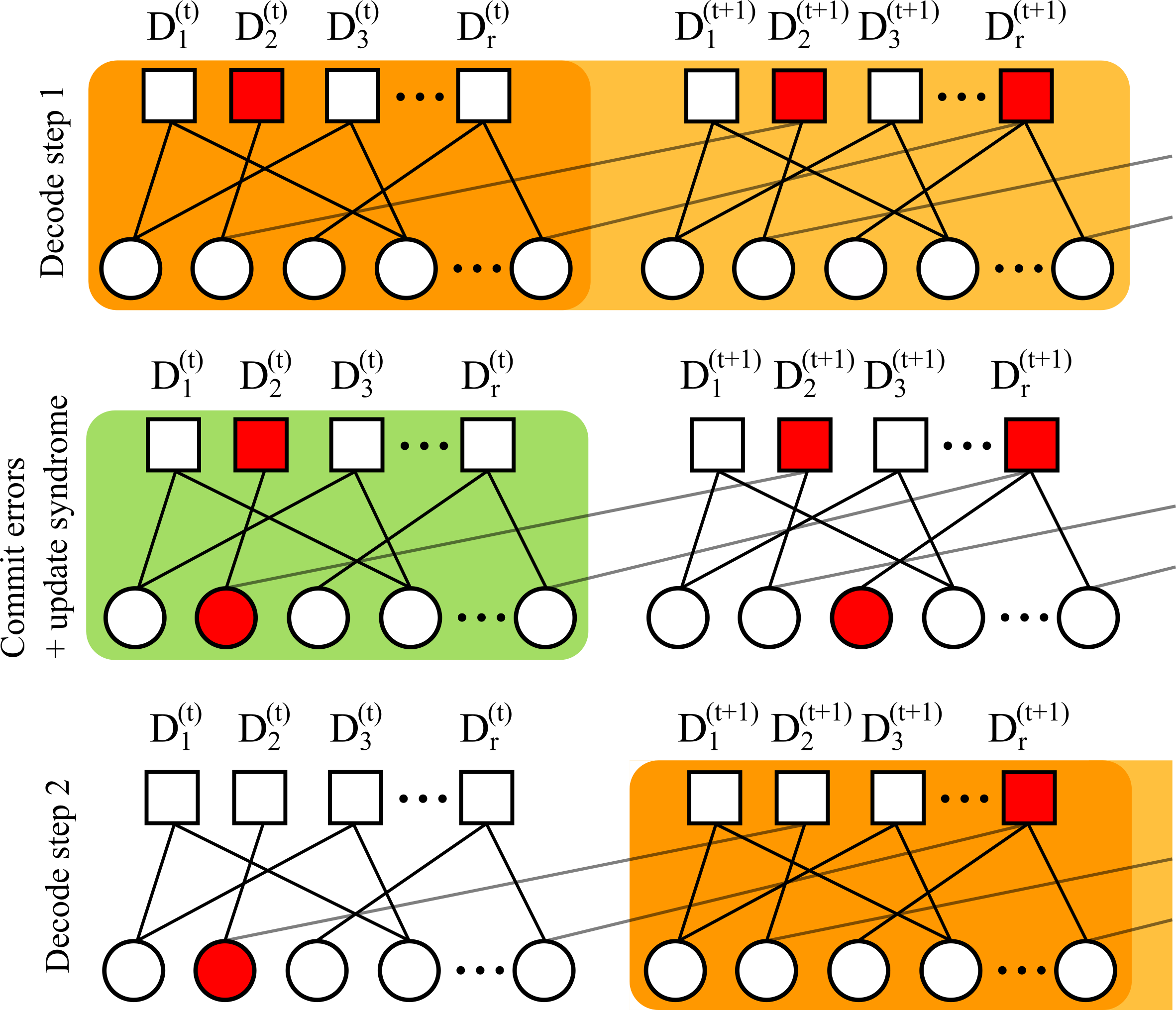}
    \caption{Schematic for a (2,1)-overlapping window decoder. Square nodes represent detectors, circle nodes represent errors, and nodes are colored red if detector is violated or an error is present, respectively. In the top panel, the syndrome information from two rounds is given as input to a decoder such as BP+OSD. The decoder outputs guesses for errors spatially and temporally located in the first two QEC rounds. Since the commit window size is one, only the error in the first round is committed. In the bottom panel, the window then slides up one round, and the process is repeated until the entire graph has been decoded. }
    \label{fig:sliding-window}
\end{figure}

While we could decode over the entire syndrome volume in this way, single-shot codes require only a single round of syndromes to be fault-tolerant to both data and measurement errors. Thus we should break up the DEM into regions which are decoded separately. One way of doing this is with an overlapping window decoder~\cite{huang2023improved, gong2024lowlatencyiterativedecodingqldpc}. 
An overlapping window decoder denoted as $(w,c)$ is defined by two parameters: $w$, the window size, and $c$ the commit size. $w$ rounds of syndromes are used as input to the decoder, which outputs a potential error over the entire $w$-size window; however, only errors temporally located within the first $c$ syndrome extraction rounds are committed to and applied as corrections.
The correction from the previous commit window is used to update the detectors for future windows, the decoding window advances $c$ rounds, and the process repeats.
See Fig.~\ref{fig:sliding-window} for a depiction of the decoding process for a $(w,c)=(2,1)$-overlapping window. There, square nodes represent detectors, circle nodes represent errors, and nodes are colored red if detector is violated or an error is present, respectively.
Unless otherwise stated, we use a $(1,1)$-overlapping window with BP+OSD, which consists of a maximimum of 30 iterations of the product-sum BP decoder followed by order-10 combination-sweep OSD.

\section{Hardware experiments}
\label{sec:experiments}

The experiments described in this section were performed on the H2 trapped-ion quantum computer~\cite{moses2023}, which, at the time of writing, was configured with 56 physical $\ce{^{171}Yb+}$ ion qubits. H2 is based on the QCCD architecture~\cite{Wineland98,Pino2021}, endowing the qubits with all-to-all connectivity through ion-transport operations. This nonlocal connectivity enables the implementation of the 4D surface code and other nonlocal codes~\cite{hong2024entangling,burton2024genons} without the use of costly circuit-level swapping. The connectivity, together with qubit reuse~\cite{Decross2023}, allows us to reuse ancilla qubits for multiple syndrome measurements within a single QEC cycle, which was crucial to running these experiments within the qubit budget of H2.

At the time of the experiments in this work, the physical layer error budget is dominated by two-qubit gate errors $\leq 1.3\times10^{-3}$, single-qubit gate errors $\sim3\times10^{-5}$, SPAM errors $\sim 1.5\times10^{-3}$, and additional memory error~\cite{DeCross2024}. Memory error is circuit dependent in the QCCD architecture since different algorithms call for different ion-transport sequences. We did not fully characterize the memory error for the specific circuits in this work, but see Ref.~\cite{DeCross2024} for a full characterization of memory errors incurred in random circuit sampling experiments with 56 qubits on H2.

\subsection{Memory and postselection}

For memory experiments, we prepare the code in the $\ket{0}_L$ or $\ket{+}_L$ logical states, perform a variable number of syndrome extraction rounds, and then destructively measure all of the data qubits. Using the final measurement results, we can construct a final noiseless syndrome as well as the value of the corresponding logical observable $\overline{Z}$ or $\overline{X}$. We then attempt to predict the value of the logical observable using the syndrome information, as described in Section~\ref{sec:decoding}. As opposed to performing the corrections in real-time~\cite{RyanAnderson2021}, this process is done after the entire shot has been completed; however, it would be feasible to implement BP+OSD in a hybrid compute environment~\cite{moses2023, dasilva2024demonstrationlogicalqubitsrepeated} on the device and then track the updates in real time.
Along with 33 data qubits, the 4D surface code has 20 $Z$-type and 20 $X$-type checks, which means we would exceed the qubit budget of H2 should we dedicate an ancilla qubit to each check. Instead, we reuse ancilla qubits and measure every check of a single type at a time, hence using 53 of the 56 available qubits.
The syndrome extraction circuits were determined to be immune to \textit{hook errors}~\cite{Dennis_2002}. Flag qubits or other more involved syndrome extraction methods were therefore not required, and instead bare ancilla syndrome extraction is used.

\begin{figure}[t!]
    \centering
    \includegraphics[width=\linewidth]{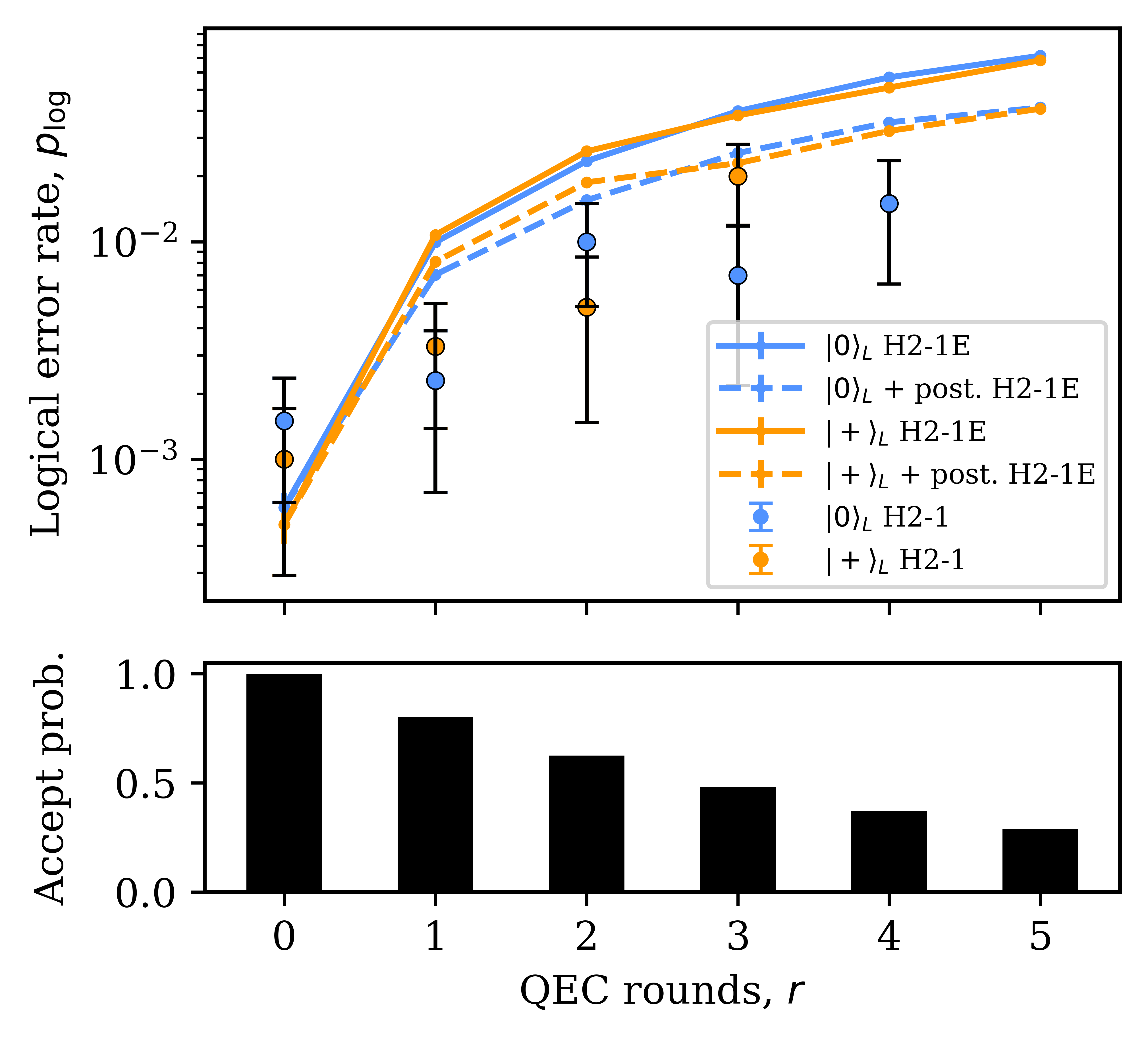}
    \caption{Memory experiments for the 4D surface code when prepared in the $\ket{0}$ and $\ket{+}$ bases and performing $r$ rounds of error correction. Solid and dashed lines represent simulated data taken from an in-house emulator with access to device error parameters and accurate ion transport scheduling. Data points with error bars correspond to circuits ran on the H2 quantum computer. The bottom bar chart indicates the probably of a shot from the emulator making it past postselection, i.e., all observed syndromes are valid. When postselecting, the resulting logical error rates are shown by the dashed lines. }
    \label{fig:memory}
\end{figure}

\begin{figure*}[t]
    \centering
    \includegraphics[width=\linewidth]{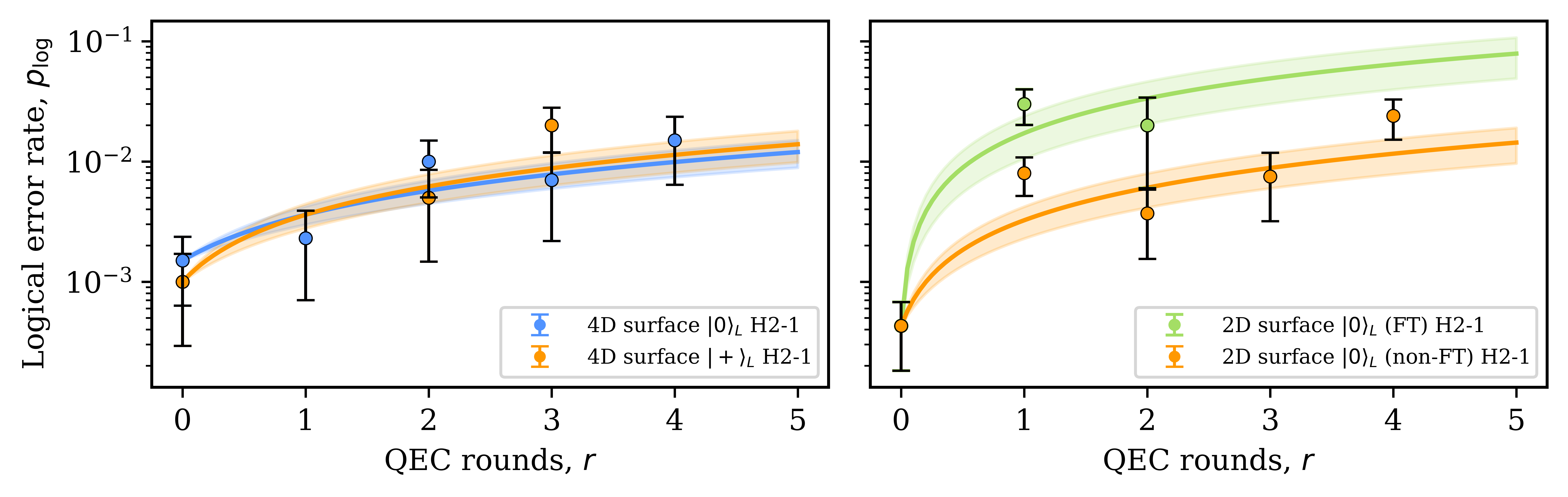}
    \caption{Fits and uncertainty regions of the logical QEC cycle error rate, $p_\text{cycle}$. To extract $p_\text{cycle}$, we perform a fit of the exponential decay curve, $p_{\log}(r) = 0.5 + (p_{\text{spam}} - 0.5)(1 - 2\cdot p_{\text{cycle}})^r$. Here $p_{\log}(r)$ is the logical error rate observed after $r$ QEC cycles, and $p_\text{spam}$ is the logical error rate observed for zero QEC cycles. The values of $p_\text{spam}$ and $p_\text{cycle}$ along with their uncertainties are also reported in Table~\ref{table:results}.}
    \label{fig:curve_fit}
\end{figure*}

Fig.~\ref{fig:memory} reports the resulting logical error rates from H2 (H2-1) and stabilizer simulations of the device (H2-1E) after performing $r$ rounds of syndrome extraction and then decoding using BP+OSD and a $(1,1)$-overlapping window.
For the $0r, ..., 4r$ circuits, we performed $\{2000,900,400,300,200\}$ shots, respectively, for each basis.
Error bars on the hardware data points are calculated using $\sigma = \sqrt{p_{\log}(1-p_{\log})/N}$ , where $N$ is the number of collected samples, and $p_{\log}$ is the logical error rate.
The simulations, indicated by solid and dashed lines, are performed with an in-house emulator utilizing a stabilizer simulator and error parameters obtained from independent measurements on the device, and they include simulating the ion-transport and memory errors. Following Ref.~\cite{RyanAnderson2021}, we estimate a QEC cycle fidelity by fitting to an exponential decay function 
\begin{equation}
    \label{eq:decay}
    p_{\log}(r) = 0.5 + (p_{\text{spam}} - 0.5)(1 - 2\cdot p_{\text{cycle}})^r.
\end{equation} 
Here $p_{\log}(r)$ is the logical error rate observed after $r$ QEC cycles and $p_{\text{spam}}$, the logical SPAM error rate, is the observed logical error rate at zero QEC cycles. 
For the 4D surface code implemented on H2, we observe a logical SPAM error rate of $1.5 \times 10^{-3} \pm 8.7 \times 10^{-4}$ when prepared in the $\ket{0}_L$ basis and $1.0 \times 10^{-3} \pm 7.1 \times 10^{-4}$ when prepared in the $\ket{+}_L$ basis. Performing a fit of Eq.~\eqref{eq:decay} to extract $p_{\text{cycle}}$ yields a logical QEC cycle error rate of $2.1\times 10^{-3}\pm 6.2\times 10^{-4}$ for the $\ket{0}_L$ basis and $2.6\times 10^{-3} \pm 8.3\times 10^{-4}$ for the $\ket{+}_L$ basis. For clarity, these values along with their uncertainties are listed in Table~\ref{table:results}. Additionally, Fig.~\ref{fig:curve_fit} displays the fits and uncertainties of Eq~\eqref{eq:decay} along with the machine data.

{
\setlength{\tabcolsep}{0.5em} % for the horizontal padding
\renewcommand{\arraystretch}{1.5}% for the vertical padding
\begin{table*}[]
\begin{tabular}{|c|c|c|c|} \hline
Code                & Basis       & $p_{\text{spam}}$                           & $p_{\text{cycle}}$                        \\ \hline
4D Surface          & $\ket{+}_L$ & $1.0 \times 10^{-3} \pm 7.1 \times 10^{-4}$ & $2.6\times 10^{-3} \pm 8.3\times 10^{-4}$ \\ \hline
4D Surface          & $\ket{0}_L$ & $1.5 \times 10^{-3} \pm 8.7 \times 10^{-4}$ & \boldmath $2.1\times 10^{-3}\pm 6.2\times 10^{-4}$ \\ \hline
2D Surface (non-FT) & $\ket{0}_L$ & \boldmath $4.3 \times 10^{-4} \pm 2.5 \times 10^{-4}$ & $2.8\times 10^{-3} \pm 9.6\times 10^{-4}$ \\ \hline
2D Surface (FT)     & $\ket{0}_L$ & \boldmath $4.3 \times 10^{-4} \pm 2.5 \times 10^{-4}$ & $1.7 \times 10^{-2} \pm 6.7 \times 10^{-3}$ \\ \hline
\end{tabular}
\caption{Observed logical SPAM and logical QEC cycle error rates from performing memory experiments on the H2 quantum computer. Logical SPAM error rates correspond to the logical error rate at zero QEC cycles. Logical QEC cycle error rates were obtained by taking the logical error rates from several $r$'s and fitting Eq.~\eqref{eq:decay}. Bolded values indicate the best performance out of the codes we investigated. }
\label{table:results}
\end{table*}
}

\begin{figure*}[t]
\centering
% \begin{minipage}[c]{\textwidth}
\centering
    \includegraphics[width=\linewidth]{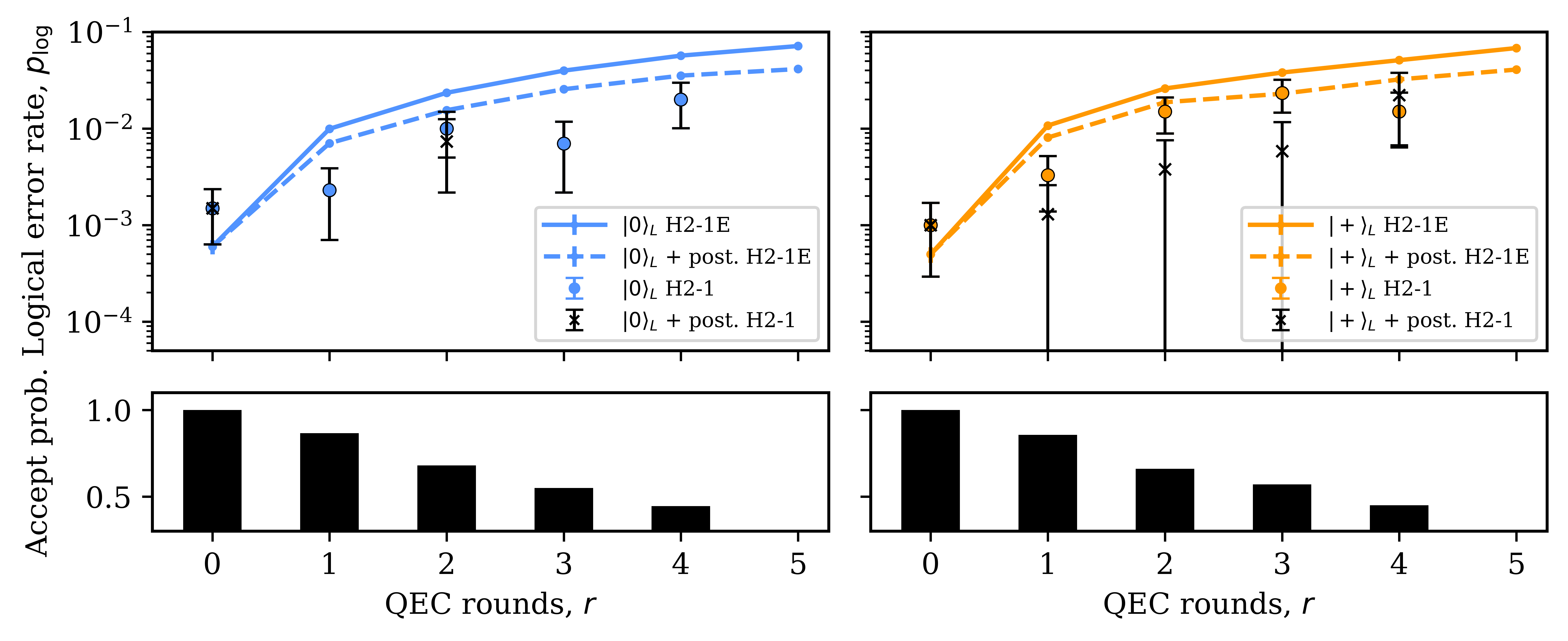}
    \caption{Memory experiments for the 4D surface code for when prepared in the $\ket{0}$ (left panels) and $\ket{+}$ (right panels) bases. Solid and dashed lines represent emulator data with no postselection and postselection on valid syndromes, respectively. Circle and X markers represent H2 hardware data with no postselection and postselection on valid syndromes, respectively. The bar charts represent the percentage of shots coming from hardware which had valid syndromes. By postselecting on the hardware data, several experiments had zero ($1r$, $3r$, $4r$ for $\ket{0}_L$) or one logical error, hence the large error bars. Nearly every experiment sees a modest improvement in the logical error rate. }
    \label{fig:memory2}
% \end{minipage}
\end{figure*}

Notably, coherent $Z$-type memory errors are not accurately captured in the stabilizer simulation. The emulator partially captures this noise by allowing memory error to accumulate coherently during idling and transport operations, but before a gating operation, the noise is projected onto a stochastic Pauli noise process to fit within the framework of the stabilizer simulation. In other words, in the simulation, memory error coherently accumulates between gates, but is not allowed to coherently propagate through gates.
Additionally, during the preparation of this paper the H2 laser systems underwent performance upgrades. These upgrades have likely improved the two-qubit gate fidelity, but a full characterization has not been carried out yet, and therefore has not been incorporated in the error parameters of the emulator. More importantly, numerical experiments indicated that the two-qubit gate fidelity was not the limiting noise source in the experiment, and that memory error dominates the logical error budget. This likely contributes to the discrepancy between our emulator and experimental results as we have made several improvements to ion-transport and cooling routines, resulting in an $\mathcal{O}(10\%)$ reduction in the typical circuit times which is currently not reflected in the emulator.

The binary linear code formed by the metachecks of the $[[33,1,4]]$ 4D surface code only has distance $d=2$, and as such we are not able to correct invalid syndromes before decoding. Instead, we can employ error detection and postselection, where we only decode shots if the metasyndrome is zero. The result of this is shown by the dashed lines in Fig.~\ref{fig:memory}, where we see a modest improvement in the logical error rate; however, this comes at the cost of discarding a significant percentage of shots. Applying postselection to the hardware data resulted in an improvement in logical fidelity for nearly every point, with some points now seeing no logical errors (see Fig.~\ref{fig:memory2}).
Scaling up the code to larger $L$ would provide the ability to correct for syndrome errors before decoding.

\subsection{Single-shot advantages}
\label{sec:advantages}

\begin{figure}[t]
    \centering
    \includegraphics[width=\linewidth]{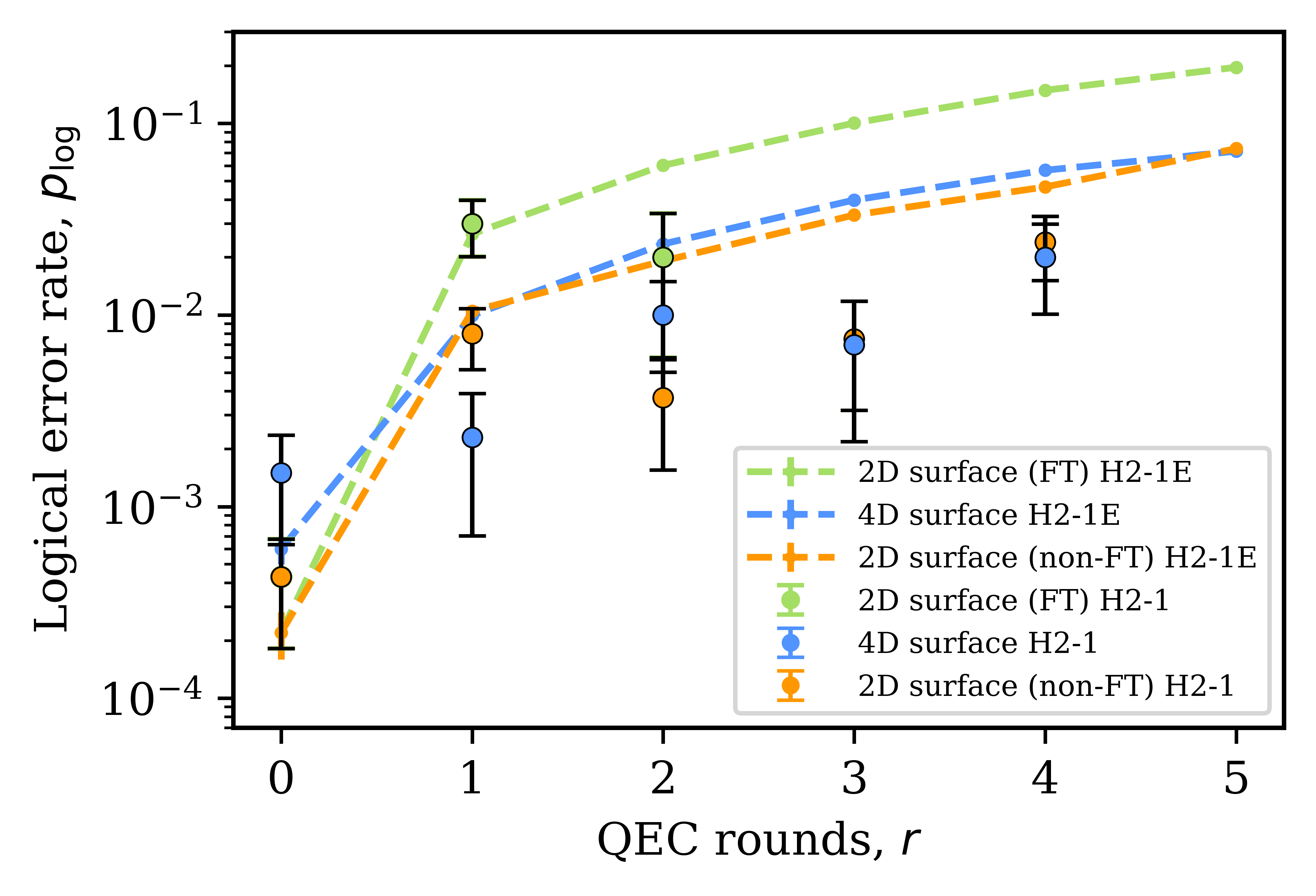}
    \caption{Semilog plot of logical error rate as a function of number of QEC cycles. The non-FT 2D surface code data consists of $r$ rounds of error correction before decoding with BP+OSD and a (1,1)-overlapping window. The FT 2D surface code data consists of $4r$ rounds of error correction before decoding with BP+OSD and a (4,4)-overlapping window. The 4D surface code data is the same as shown in Fig.~\ref{fig:memory}. }
    \label{fig:2dvs4d}
\end{figure}

As mentioned in Section~\ref{sec:single-shot}, one benefit of using codes that allow for single-shot error correction is that syndrome extraction does not need to be repeated in order to be fault-tolerant. Consequently, this has the potential of decreasing the depth of the syndrome extraction circuit by $\Theta(d)$. The drawback is that often larger codeblocks are required for the code to be single-shot, leading to deeper syndrome extraction circuits in general.

As a comparison, we look at the $[[25,1,4]]$ surface code. We can assign a physical qubit to each generator for a total of 49 qubits, which fits on the 56 qubit H2 device. To replicate the conditions of the 4D surface code, we instead reuse ancilla qubits (for a total of 37 physical qubits) and measure all checks of a single type first, then measure all checks of the other type.
Disregarding initialization circuits, a single round of syndrome extraction for the $L=2$ 4D surface code requires 168 CNOT gates. This is in contrast to a single round of syndrome extraction for the $L=4$ 2D surface code which requires only 84 CNOT gates. 
However, as this code is not single-shot, we need to do $d=4$ syndrome extraction rounds to be fault-tolerant to both data and measurement errors, hence requiring 336 CNOT gates. 
Note that with sufficient parallelization and connectivity, the circuit depth is constant at depth-4 for the 2D surface code~\cite{Tomita_2014} and depth-8 for the 4D surface code~\cite{Duivenvoorden_2019}.
Given that the H2 device operates with four gate zones, full parallelization cannot be achieved. 

We can estimate the asymptotic gate count scaling as follows. For a lattice of size $L$, the parameters of the 2D and 4D surface codes, respectively, are $[[O(L^2),1,L]]$ and $[[O(L^4),1,L^2]]$. The stabilizer generators have weight bounded by 4 and 6, respectively. Since the 4D surface code is single-shot, we can therefore expect to need $O(L^4)$ CNOT gates for each syndrome extraction. Alternatively, when we construct a 2D surface code with $d = L^2$ we get FT syndrome extraction circuits requiring $O(L^4)\cdot O(L^2) = O(L^6)$ CNOT gates, where the $O(L^2)$ factor comes from the repeated measurements.
This analysis also holds for the slightly more qubit efficient rotated surface code.

In Fig.~\ref{fig:2dvs4d}, we show the results of memory experiments using the FT and non-FT syndrome extraction circuits for the $[[25,1,4]]$ 2D surface code. 
Since we are restricting ourselves to fewer ancilla qubits than is necessary to perform the depth-4 syndrome extraction circuit, we instead measure all generators of a single type followed by all of the opposite type generators. 
The non-FT data consists of $r$ rounds of syndrome extraction and is decoded using BP+OSD and a (1,1)-overlapping window. The FT data consists of $4r$ rounds of syndrome extraction and is decoded using a (4,4)-overlapping window to represent taking $d=4$ rounds of syndrome data before attempting to decode. The final syndrome is still used when decoding the FT data; however, since it is assumed to be noiseless, we do not need repeated measurements and can instead use a $(1,1)$-overlapping window to provide the final correction. For the $0r, ..., 4r, 8r$ circuits, we performed $\{7000,1000,800,400,300, 100\}$ shots, respectively. Here, the $8r$ circuit either represents eight non-FT rounds or two FT rounds, depending on choice of window when decoding.

Despite the 4D surface using more qubits and requiring a syndrome extraction circuit twice as long as the non-FT 2D surface code, the emulated performance between them is nearly identical after initialization. The hardware data, while less clear, shows comparable performance between the two as well. See Table~\ref{table:results} for the 2D surface code data displayed alongside the 4D surface code data.
For the 2D surface code prepared in the $\ket{0}_L$ basis on H2, we measure an average logical SPAM error of $4.3 \times 10^{-4} \pm 2.5 \times 10^{-4}$, which is significantly better than the corresponding 4D surface code error rate.
However, this difference disappears when performing additional QEC cycles, indicating that the 2D surface code suffers in the single-shot regime, while the 4D surface code is more robust. This is corroborated by fitting Eq.~\eqref{eq:decay} and extracting a hardware logical QEC cycle error rate of $2.8\times 10^{-3} \pm 9.6\times 10^{-4}$ for the 2D surface code.
We still see the worst performance out of the FT 2D surface code, where the increased circuit depth is not offset by the $\Theta(d)$ repeated syndrome measurements: we observe a logical QEC cycle error rate of $1.7 \times 10^{-2}  \pm 6.7 \times 10^{-3}$, nearly an order of magnitude worse than the 4D and the non-FT 2D surface codes.

\section{Discussion}
\label{sec:discussion}

In this work, we have implemented the [[33,1,4]] 4D surface code and performed the first experimental demonstration of single-shot quantum error correction with bare ancilla qubits. When compared to the 2D surface code in memory experiments, the 4D surface code performs nearly identically in the single-shot regime: the 2D surface code achieves a hardware logical QEC cycle error rate of $2.8\times 10^{-3} \pm 9.6\times 10^{-4}$ compared to the logical QEC cycle error rate achieved by the 4D surface code, $2.1\times 10^{-3}\pm 6.2\times 10^{-4}$.
This is despite the 4D surface code using 16 more qubits and requiring a syndrome extraction circuit with twice as many CNOTs as the 2D surface code. We also note that when comparing FT implementations for the two codes, the 4D code outperforms the 2D code by nearly an order of magnitude ($2.1\times 10^{-3}\pm 6.2\times 10^{-4}$ versus $1.7 \times 10^{-2}  \pm 6.7 \times 10^{-3}$). This fidelity gain is in addition to the asymptotic gain in wall-clock time associated with single-shot error correction. We note that as device error rates improve and larger codes are implemented, the outperformance of the non-FT implementation is unlikely to hold. Indeed, to be guaranteed a threshold~\cite{Aharonov_Ben-Or_1999}, FT QEC is required. 
In the large-scale regime, we thus expect the 4D surface code and the FT 2D surface code to perform similarly; however, as noted Sec.~\ref{sec:advantages} the 4D surface code still provides a reduction in the number of CNOT gates.

We showed that single-shot codes may provide advantages in terms of performance, circuit depth, and clock-speed. Clock-speed is a notable issue for ion-trap quantum computers, which have gate times on the order of microseconds, compared to nanosecond gate times of superconducting devices. Developing ways to speed up fault-tolerant quantum computations will help provide competitive time to solution.
In addition to using single-shot codes, reductions in the QEC cycle time may be achieved by applying partial syndrome QEC~\cite{Berthusen_2024} or algorithmic fault-tolerance~\cite{zhou2024algorithmicfaulttolerancefast}, among other theoretical and hardware improvements. 

The ultimate code of choice for QCCD quantum computers is far from certain. While the 4D surface code may not be the best option for large-scale quantum computation due to its poor encoding rate, codes based on the repeated homological product of good classical codes~\cite{Campbell_2019, Higgott_2023} may provide the same benefits at more favorable rates; however, their large blocklengths make them poorly suited for near-term hardware experiments. 
Given the all-to-all connectivity inherent to the QCCD architecture, nonlocal quantum LDPC codes are an appealing option which may provide significant time and space overhead reductions as compared to topological codes.

\section*{Acknowledgements}

We acknowledge helpful discussions with Ciar\'an Ryan-Anderson, Natalie Brown, Elijah Durso-Sabina, and Shival Dasu. 
We thank the entire hardware team at Quantinuum for making these experiments possible.

% \newpage

\bibliography{bibliography}

\appendix

\clearpage

\section{Code construction}
\label{apx:code_construction}

In this appendix, we present an explicit construction for the $D$-dimensional surface code in the framework of chain complexes and $\mathbb{F}_2$-homology. 

To construct a quantum code we take the homological~\cite{Bravyi2014}, or hypergraph~\cite{Tillich_2014} product of two 2-term chain complexes. The difference between the two products is subtle and concerns the use of a transpose Tanner graph, $D$ in Eq.~\eqref{eq:base_complexes}. The construction described here corresponds to the hypergraph product. The first step of the product is to take the \textit{double complex} of two chain complexes, $C \boxtimes D$, which is equipped with vertical boundary maps $\partial_i^v = \partial_i^C \otimes \mathbb{I}_{D_i}$ and horizontal boundary maps $\partial_i^h = \mathbb{I}_{C_i} \otimes \partial_i^D$. 
Here we use the notation $\partial_i^A$ to denote the $i$th boundary operator of the chain complex $A$.
See Fig.~\ref{fig:2dcommdiagram} for an example of a double complex arising from the tensor product of two 2-term chain complexes.
From a double complex, we associate the \textit{total complex} by performing the direct sum over vector spaces and boundary maps of like dimension
\begin{align}
    \label{eq:tot_complex}
    \text{Tot}(C \boxtimes D)_i &= \bigoplus_{i=j+k} C_j \otimes D_k = E_i \\
    \partial_i^E &= \bigoplus_{i=j+k} \partial_j^v \oplus \partial_k^h.
\end{align}
The resulting chain complex, deemed the \textit{tensor product} of $C$ and $D$, $C \otimes D$, can be used to construct a CSS code by choosing some consecutive three-term sequence. The parameters of the resulting code can be explicitly calculated or the K\"unneth formula can be applied,
\begin{equation}
    \label{eq:kunneth}
    H_i(C \otimes D) \cong \bigoplus_{i=j+k} H_j(C) \otimes H_k(D).
\end{equation}
While the distance of a code is in general difficult to calculate, it was shown in Ref.~\cite{Zeng_2019} that the distances of a tensor product of an arbitrary-length chain complex with a 2-term chain complex can be calculated exactly,
\begin{equation}
    \label{eq:distance}
    d_i(C \otimes D) = \min\big(d_i(C)d_0(D), d_{i-1}(C)d_1(D)\big).
\end{equation}

Now, consider the following $L-1 \times L$ parity check matrix,
\begin{equation}
\renewcommand\arraystretch{1}
    H = \begin{pmatrix} 
    1 & 1 & &&&  \\
     & 1 & 1 &  &&& \\
     & & & \ddots && & \\
     &  &     &&1 & 1 
    \end{pmatrix}
\end{equation}
corresponding to the repetition code with parameters $[L,1,L]$. We define the following two 2-term chain complexes which will be the basis for the product constructions,
\begin{equation}
    \label{eq:base_complexes}
    \begin{tikzcd}[every matrix/.append style={},
    every label/.append style = {font = \footnotesize}]
        C = C_1 \arrow[r, "\partial"] & C_0
    \end{tikzcd} \qquad
    \begin{tikzcd}[every matrix/.append style={},
    every label/.append style = {font = \footnotesize}]
        D = D_1 \arrow[r, "\partial^T"] & D_0
    \end{tikzcd}
\end{equation}
with $\partial = H$ and $\partial^T = H^T$. Note that the procedure described below can be performed with the $L \times L$ repetition code and only the chain complex $C$, in which case one would obtain the $D$-dimensional toric code. 

\subsection{2D surface code}
To obtain the 2D surface code, we take the tensor product of $C$ and $D$, yielding the double complex as shown in Fig.~\ref{fig:2dcommdiagram}.
Performing a direct sum over vector spaces of equal dimension according to Eq.~\eqref{eq:tot_complex} results in the tensor product complex 
\begin{figure}[!t]
    \centering
    \begin{tikzcd}[every matrix/.append style={},
    every label/.append style = {font = \footnotesize},
    row sep=large, column sep=large]
        C_1 \otimes D_1 \arrow[r, "\mathbb{I}_{C_1} \otimes \partial^T"] \arrow[d, "\partial \otimes \mathbb{I}_{D_1}"] & C_1 \otimes D_0 \arrow[d, "\partial \otimes \mathbb{I}_{D_0}"] \\
        C_0 \otimes D_1 \arrow[r, "\mathbb{I}_{C_0} \otimes \partial^T"] & C_0 \otimes D_0
    \end{tikzcd}
    \caption{A commutative diagram representing the double complex that arises from the tensor product of two 2-term chain complexes.}
    \label{fig:2dcommdiagram}
\end{figure}
% [every matrix/.append style={nodes={font=\large}},
%     every label/.append style = {font = \small},
%     row sep=large, column sep=small]
\begin{equation}
    \label{eq:2d_surface_chain}
    \begin{tikzcd}[every matrix/.append style={},
    every label/.append style = {font = \footnotesize},
    row sep=large, column sep=small]
        C_1 \otimes D_1 \arrow[r, "\partial_2"] & C_0 \otimes D_1 \oplus C_1 \otimes D_0 \arrow[r, "\partial_1"] & C_0 \otimes D_0
    \end{tikzcd}
\end{equation}
where
\begin{align}
    \partial_2 &= \begin{pmatrix} 
    \partial \otimes \mathbb{I}_{D_1} \\
    \mathbb{I}_{C_1} \otimes \partial^T 
    \end{pmatrix} \label{eq:2dpartial2}  \\
    \partial_1 &= \begin{pmatrix}
        \mathbb{I}_{C_0} \otimes \partial^T & \partial \otimes \mathbb{I}_{D_0}
    \end{pmatrix}. \label{eq:2dpartial1}
\end{align}
It can be easily verified that $\partial_1\partial_2 = 0$, and so Eq.~\eqref{eq:2d_surface_chain} is a valid chain complex. Let us relabel the vector spaces and boundary maps in Eq.~\eqref{eq:2d_surface_chain} to 
$\begin{tikzcd}[every matrix/.append style={},
    every label/.append style = {font = \footnotesize},
    row sep=large, column sep=small]
        E = E_2 \arrow[r, "\partial_2^E"] & E_1 \arrow[r, "\partial_1^E"] & E_0.
\end{tikzcd}$ 
For each vector space $E_i$, we calculate the dimension $\dim E_i$.
\begin{align}
    \dim E_2 &= \dim (\mathbb{F}_2^L \otimes \mathbb{F}_2^{L-1}) = L(L-1) \\ 
    \dim E_1 &= \dim (\mathbb{F}_2^{L-1} \otimes \mathbb{F}_2^{L-1}) \\
             &+ \dim (\mathbb{F}_2^{L} \otimes \mathbb{F}_2^L) = L^2 + (L-1)^2 \nonumber \\
    \dim E_0 &= \dim (\mathbb{F}_2^{L-1} \otimes \mathbb{F}_2^{L}) = (L-1)L
\end{align}

To consider $E$ a CSS code, we identify the $L^2 + (L-1)^2$ physical qubits with 1-chains. Parity check matrices are then assigned to the boundary operators $\partial_2^E = H_Z^T$ and $\partial_1^E = H_X$. We can determine the number of logical qubits by calculating the dimension of the first homology group, $\dim H_1(E)$. 
Using the equivalence $\text{rank}(A \otimes B) = \text{rank}(A)\cdot\text{rank}(B)$, we arrive at the fact that $\text{rank}(\partial_2^E) = \text{rank}(\partial_1^E) = L(L-1)$.
Using the definition of $H_1(E)$ in Eq.~\eqref{eq:homology_group}, we have that
\begin{align}
    \dim H_1(E) &= \dim \ker \partial_1^E - \dim \text{im} \ \partial_2^E \\
      &= L^2 + (L-1)^2 - 2L(L-1) = 1, \nonumber
\end{align}
giving us a single logical qubit for the 2D surface code, as expected. Alternatively, and perhaps more elegantly, we can apply the K\"unneth formula, Eq.~\eqref{eq:kunneth}.
Here, $H_1(C) \cong \mathbb{Z}_2$, and since the checks are linearly independent $H_0(C) = 0$. As the binary linear code , $H_1(D) = 0$ and $H_0(D) \cong \mathbb{Z}_2$.
Plugging into Eq.~\eqref{eq:kunneth} as expected yields, 
\begin{align}
    \dim H_1(E) &= \dim H_1(C\otimes D) \\
    = \dim \big(H_1(C) &\otimes H_0(D) \oplus H_0(C) \otimes H_1(D)\big) = 1. \nonumber
\end{align}

We can use Eq.~\eqref{eq:distance} to calculate the distances of the resulting chain complex. 
The distance $d_0$ of the homology group $H_0(C)$ is $d_0 = 1$, unless $\partial_1$ has full column-rank in which case $d_0 = \infty$.
For $i > 0$, we say $d_i = \infty$ if $H_i(C)$ is trivial. 
Otherwise, $d_i$ is the minimum weight of a non-zero vector $x \in \ker \partial_i$, i.e., the distance of binary linear code defined by $\partial_i$. Hence we obtain,
\begin{align}
    d_1(E) &= \min \big(d_1(C)\ d_0(D), d_0(C)\ d_1(D) \big) \\
                     &= \min \big(L\cdot 1, 1 \cdot \infty) = L \nonumber
\end{align}
Technically, this is only $d_Z$, and $d = \min(d_Z, d_X)$. $d_X$ can be obtained by taking the minimum Hamming weight of non-zero elements in the cohomology classes, $H^i(C)$. In this instance we obtain the same value for $d_X$. Thus the CSS code defined by Eq.~\eqref{eq:2d_surface_chain} has the parameters $[[L^2+(L-1)^2,1,L]]$.

\subsection{3D surface code}

\begin{figure}[t!]
    \centering
    \includegraphics[width=\linewidth]{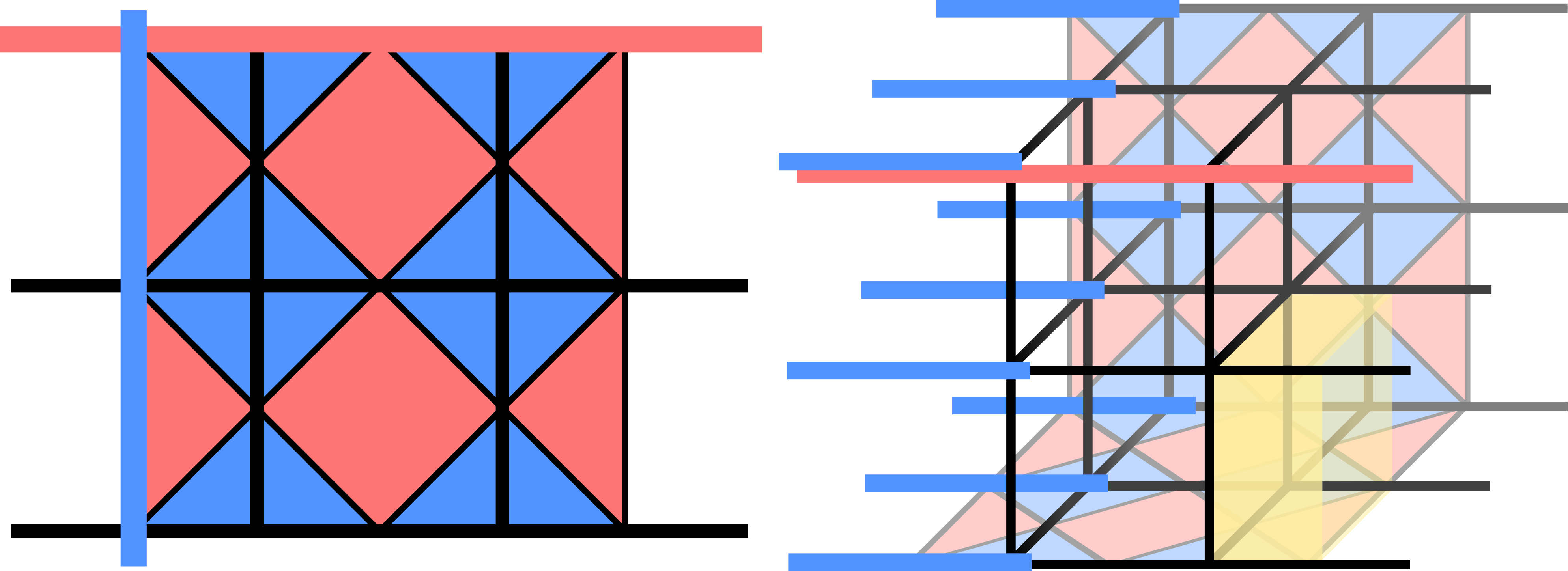}
    \caption{(a) 2D surface. Qubits are located on edges, and blue (pink) faces denote $X$ ($Z$) stabilizer generators. Blue (pink) edge highlights denote the minimum weight logical Pauli $\overline{X}$ ($\overline{Z}$). (b) 3D surface code. Qubits again are located on edges. In the 3D surface code we now have a metacheck for a single type, as indicated by the yellow box. }
    \label{fig:2d3d_codes}
\end{figure}

To construct higher dimensional surface codes, we iteratively apply the same process from which we obtained the 2D surface code. The 3D surface code can be seen as the hypergraph product of a repetition code with the 2D surface code, so we take the tensor product of $E$ and $D$. Fig.~\ref{fig:3dcommdiag} shows the resulting double complex, from which we again collect like-dimension vector spaces and obtain a new 4-term chain complex
\begin{figure}[!t]
    \centering
    \begin{tikzcd}[every matrix/.append style={},
    every label/.append style = {font = \footnotesize},
    row sep=large, column sep=large]
        E_2 \otimes D_1 \arrow[r, "\mathbb{I}_{E_2} \otimes \partial^T"] \arrow[d, "\partial_2^E \otimes \mathbb{I}_{D_1}"] & E_2 \otimes D_0 \arrow[d, "\partial_2^E \otimes \mathbb{I}_{D_0}"] \\
        E_1 \otimes D_1 \arrow[r, "\mathbb{I}_{E_1} \otimes \partial^T"] \arrow[d, "\partial_1^E \otimes \mathbb{I}_{D_1}"] & E_1 \otimes D_0 \arrow[d, "\partial_1^E \otimes \mathbb{I}_{D_0}"] \\
        E_0 \otimes D_1 \arrow[r, "\mathbb{I}_{E_0} \otimes \partial^T"] & E_0 \otimes D_0
    \end{tikzcd}
    \caption{A commutative diagram representing the double complex that arises from the tensor product of 2-term and a 3-term chain complex.}
    \label{fig:3dcommdiag}
\end{figure}
\begin{equation}
    \label{eq:3d_chain}
    \begin{tikzcd}[every matrix/.append style={},
    every label/.append style = {font = \footnotesize},
    row sep=large, column sep=small]
        F = F_3 \arrow[r, "\partial_3^F"] & F_2 \arrow[r, "\partial_2^F"] & F_1 \arrow[r, "\partial_1^F"] & F_0
    \end{tikzcd}
\end{equation}
with boundary maps
\begin{align}
    \partial_3^F &= \begin{pmatrix} 
    \partial_2^E \otimes \mathbb{I}_{D_1}  \\
    \mathbb{I}_{E_2} \otimes \partial^T 
    \end{pmatrix} \\
    \partial_2^F &= \begin{pmatrix} 
    \partial_1^E \otimes \mathbb{I}_{D_1} & 0  \\
    \mathbb{I}_{E_1} \otimes \partial^T & \partial_2^E \otimes \mathbb{I}_{D_0}
    \end{pmatrix} \\
    \partial_1^F &= \begin{pmatrix}
        \mathbb{I}_{E_0} \otimes \partial^T & \partial_1^E \otimes \mathbb{I}_{D_0}
    \end{pmatrix}.
\end{align}

We now have a choice as to which vector space we identify with qubits, $F_2$ or $F_1$. The choice corresponds to determining which logical type is planar-like (with weight $L^2$) versus string like (with weight $L$). 
Whichever logical is planar-like, the same type stabilizers have \text{metachecks} (see Fig.~\ref{fig:2d3d_codes}(b)). Refer to Sec.~\ref{sec:single-shot} on metachecks and single-shot codes.
For instance, let $\partial_2^F = H_Z^T$ and $\partial_1^F = H_X$, in which case we identify qubits with vectors in $F_1$. We then have the dimensions
\begin{align}
    % \dim F_3 &= \dim (E_2 \otimes D_1) = L(L-1)^2 \\
    \dim F_2 &= \dim (E_1 \otimes D_1 \oplus E_2 \otimes D_0) \\
             &= 2L^2(L-1) + (L-1)^3 \nonumber \\
    \dim F_1 &= \dim (E_0 \otimes D_1 \oplus E_1 \otimes D_0) \\
             % &= L^3 + L(L-1)^2 + L(L-1)^2 = 
             &= L^3 + 2L(L-1)^2 \nonumber. \\
    % \dim F_0 &= \dim (E_0 \otimes D_0) = L^2(L-1).
\end{align}
With this choice of qubits, we have $Z$-type metachecks where the metacheck matrix is given by $\partial_3^F = M_Z^T$. Going through the above process again to calculate the dimension of the homology group $H_1(F) = H_1(E)\otimes H_0(D) \oplus H_0(E)\otimes H_1(D)$ and the corresponding distances $d_1(F), d^1(F)$, yields a quantum CSS code with parameters $[[L^3 + 2L(L-1)^2, 1, ~\min(L, L^2)]]$.

\subsection{4D surface code}
\label{sec:4dsurfaceconstruction}
\begin{figure}[]
    \centering
    \begin{tikzcd}[every matrix/.append style={},
    every label/.append style = {font = \footnotesize},
    row sep=large, column sep=large]
        F_3 \otimes C_1 \arrow[r, "\mathbb{I}_{F_3} \otimes \partial"] \arrow[d, "\partial_3^F \otimes \mathbb{I}_{C_1}"] & F_3 \otimes C_0 \arrow[d, "\partial_3^F \otimes \mathbb{I}_{C_0}"] \\
        F_2 \otimes C_1 \arrow[r, "\mathbb{I}_{F_2} \otimes \partial"] \arrow[d, "\partial_2^F \otimes \mathbb{I}_{C_1}"] & F_2 \otimes C_0 \arrow[d, "\partial_2^F \otimes \mathbb{I}_{C_0}"] \\
        F_1 \otimes C_1 \arrow[r, "\mathbb{I}_{F_1} \otimes \partial"] \arrow[d, "\partial_1^F \otimes \mathbb{I}_{C_1}"] & F_1 \otimes C_0 \arrow[d, "\partial_1^F \otimes \mathbb{I}_{C_0}"] \\
        F_0 \otimes C_1 \arrow[r, "\mathbb{I}_{F_0} \otimes \partial"] & F_0 \otimes C_0 
    \end{tikzcd}
    \caption{A commutative diagram representing the double complex that arises from the tensor product of 2-term and a 4-term chain complex.}
    \label{fig:4dcommdiag}
\end{figure}

Finally, to obtain the 4D surface code we again apply the product construction. Although this time note that we take the tensor product of $F$ and $C$, instead of $D$ as the second term in the product (see Fig.~\ref{fig:4dcommdiag}). Collecting like-dimension vector spaces yields a 5-term chain complex
\begin{equation}
    \begin{tikzcd}[every matrix/.append style={},
    every label/.append style = {font = \footnotesize},
    row sep=large, column sep=small]
        G = G_4 \arrow[r, "\partial_4^G"] & G_3 \arrow[r, "\partial_3^G"] & G_2 \arrow[r, "\partial_2^G"] & G_1 \arrow[r, "\partial_1^G"] & G_0
    \end{tikzcd}
\end{equation}
with boundary maps
\begin{align}
    \partial_4^G &= \begin{pmatrix} 
    \partial_3^F \otimes \mathbb{I}_{C_1}  \\
    \mathbb{I}_{F_3} \otimes \partial
    \end{pmatrix} \\
    \partial_3^G &= \begin{pmatrix} 
    \partial_2^F \otimes \mathbb{I}_{C_1} & 0  \\
    \mathbb{I}_{F_2} \otimes \partial & \partial_3^F \otimes \mathbb{I}_{C_0}
    \end{pmatrix} \label{eq:4dpartial1} \\
    \partial_2^G &= \begin{pmatrix} 
        \partial_1^F \otimes \mathbb{I}_{C_1} & 0  \\
        \mathbb{I}_{F_1} \otimes \partial & \partial_2^F \otimes \mathbb{I}_{C_0}
    \end{pmatrix} \label{eq:4dpartial2} \\
    \partial_1^G &= \begin{pmatrix}
        \mathbb{I}_{F_0} \otimes \partial & \partial_1^F \otimes \mathbb{I}_{C_0}
    \end{pmatrix}.
\end{align}

In order to have metachecks for both $X$ and $Z$ syndromes, we must identify qubits with elements of $G_2$. Then we have that $M_Z = \partial_4^G, H_Z^T = \partial_3^G, H_X = \partial_2^G, M_X = \partial_1^G$. % check for correct transpose.
Going through the above process yields
\begin{align}
    \dim G_2 &= \dim (F_1 \otimes C_1 \oplus F_2 \otimes C_0) \\
             &= L^4 + 4L^2(L-1)^2 + (L-1)^4 \nonumber \\
             &= 6L^4 - 12L^3 + 10L^2 - 4L + 1 \nonumber
\end{align}
The number of logical qubits is then given by $\dim H_2(G) = \dim \big( H_2(F)\otimes H_0(D) ~\oplus~ H_1(E) \otimes H_1(D) ~\oplus H_0(E)~ \otimes H_2(D) \big)$. 
Hence the code parameters are $[[6L^4 - 12L^3 + 10L^2 - 4L + 1,1,L^2]]$. For the $L=2$ case as discussed in the main text, we get a $[[33,1,4]]$ code.

Achieving even higher dimensional surface codes can be done by alternating $C$ and $D$ and taking the product with the chain complex representing the surface code one dimension lower. An equivalent method to iteratively increasing the dimension as we did here is to take the tensor product of $D$ repetition codes at once~\cite{Zeng_2019}, obtaining triple, quadruple, or higher dimensional complexes before summing and to obtain a total complex representing the $D$-dimensional surface/toric code. Although by performing the iterative tensor product, we are able to use Eq.~\eqref{eq:distance} to calculate the resulting distances.

\section{Explicit construction of the 4D surface code}

For pedagogical completeness, we go through the procedures described in Appendix~\ref{apx:code_construction} to obtain the pcms for the $L=2$ 2D, 3D, and 4D surface codes. 

For $L=2$, we obtain the $1 \times 2$ repetition code, $H = \begin{pmatrix}
1 & 1
\end{pmatrix}$ with parameters $[2,1,2]$. $H$ and $H^T$ then serve as boundary maps for the chain complexes $C$ and $D$, respectively, as shown in Eq.~\eqref{eq:base_complexes}. Applying the first iteration of the hypergraph product yields a 2D surface code with parameters $[[5,1,2]]$ and pcms given in Eqs.~\eqref{eq:2dpartial2},~\eqref{eq:2dpartial1} and shown explicitly below
\begin{align}
    H_Z^T = \partial_2^{E} = \left(\begin{array}{c}
1  1  1  0  0\\
1  0  0  1  1
\end{array}\right) \\  
H_X = \partial_1^{E} =\left(\begin{array}{c}
1  1  0  1  0\\
1  0  1  0  1
\end{array}\right)
\end{align}
As this code has distance $d=2$, it is only able to detect a single error. Note that in the main text we investigate the $[[25,1,4]]$ 2D surface code, which we can obtain by using $L=4$.

We now apply a second iteration of the hypergraph product to obtain the $L=2$ 3D surface code. Following Fig.~\ref{fig:3dcommdiag}, we take the the chain complex $D$ and the chain complex $E$ representing the $L=2$ 2D surface code and obtain a new 4-term chain complex representing the $L=2$ 3D surface code, Eq.~\eqref{eq:3d_chain}, which has parameters $[[12,1,2]]$.
Explicitly, we obtain the following pcms:

\setcounter{MaxMatrixCols}{20}
\begin{align}
    H_Z^T = \partial_2^{F} = \left(\begin{array}{c}
1 1 1 1 0 0 0 0 0 0 0 0\\ 
1 0 0 0 1 1 0 0 0 0 0 0\\
0 1 0 0 0 0 1 1 0 0 0 0\\ 
1 0 0 0 0 0 0 0 1 1 0 0\\ 
0 1 0 0 0 0 0 0 0 0 1 1\\ 
0 0 1 0 1 0 1 0 0 0 0 0\\ 
0 0 0 1 0 1 0 1 0 0 0 0\\ 
0 0 1 0 0 0 0 0 1 0 1 0\\ 
0 0 0 1 0 0 0 0 0 1 0 1
\end{array}\right) \\ 
H_X = \partial_1^{F} = \left(\begin{array}{c}
1 0 1 0 1 0 0 0 1 0 0 0\\ 
1 0 0 1 0 1 0 0 0 1 0 0\\ 
0 1 1 0 0 0 1 0 0 0 1 0\\ 
0 1 0 1 0 0 0 1 0 0 0 1
\end{array} \right)
\end{align}
For the 3D surface code, there is an asymmetry between the $Z$- and $X$-bases. Indeed, we have that the $Z$-distance $d_Z = 2$ while the $X$-distance $d_X = 4$, hence leading to the parameters $[[12,1,2]]$. In addition to the $Z$- and $X$-type pcms, we have a metacheck matrix for the $Z$-type checks. 
\begin{align}
M_Z^T = \partial_3^F = \left(\begin{array}{c}
1 1 1 0 0 1 1 0 0\\
1 0 0 1 1 0 0 1 1
\end{array}\right)
\end{align}
The classical code formed by this metacheck matrix has distance $d=2$, and so it can detect, but not correct, a single syndrome error on the $Z$-type syndromes.

We finally obtain the $L=2$ 4D surface code by taking the hypergraph product with the chain complex $C$ and the chain complex representing the 3D surface code, Eq~\eqref{eq:3d_chain}, to obtain a 5-term chain complex $G$, Eq.~\eqref{eq:4dchain}, representing the 4D surface code. This code balances out the asymmetry of the 3D surface code and has parameters $[[33,1,4]]$, with both $d_X = d_Z = 4$. 
The full pcms are shown below.

\begin{align}
   H_Z^T = \partial_3^G =  \left( \begin{array}{c}
        1 0 1 0 1 0 1 0 0 0 0 0 0 0 0 0 0 0 0 0 0 0 0 0 1 0 0 0 0 0 0 0 0\\ 
0 1 0 1 0 1 0 1 0 0 0 0 0 0 0 0 0 0 0 0 0 0 0 0 1 0 0 0 0 0 0 0 0\\ 
1 0 0 0 0 0 0 0 1 0 1 0 0 0 0 0 0 0 0 0 0 0 0 0 0 1 0 0 0 0 0 0 0\\ 
0 1 0 0 0 0 0 0 0 1 0 1 0 0 0 0 0 0 0 0 0 0 0 0 0 1 0 0 0 0 0 0 0\\ 
0 0 1 0 0 0 0 0 0 0 0 0 1 0 1 0 0 0 0 0 0 0 0 0 0 0 1 0 0 0 0 0 0\\ 
0 0 0 1 0 0 0 0 0 0 0 0 0 1 0 1 0 0 0 0 0 0 0 0 0 0 1 0 0 0 0 0 0\\ 
1 0 0 0 0 0 0 0 0 0 0 0 0 0 0 0 1 0 1 0 0 0 0 0 0 0 0 1 0 0 0 0 0\\ 
0 1 0 0 0 0 0 0 0 0 0 0 0 0 0 0 0 1 0 1 0 0 0 0 0 0 0 1 0 0 0 0 0\\ 
0 0 1 0 0 0 0 0 0 0 0 0 0 0 0 0 0 0 0 0 1 0 1 0 0 0 0 0 1 0 0 0 0\\ 
0 0 0 1 0 0 0 0 0 0 0 0 0 0 0 0 0 0 0 0 0 1 0 1 0 0 0 0 1 0 0 0 0\\ 
0 0 0 0 1 0 0 0 1 0 0 0 1 0 0 0 0 0 0 0 0 0 0 0 0 0 0 0 0 1 0 0 0\\ 
0 0 0 0 0 1 0 0 0 1 0 0 0 1 0 0 0 0 0 0 0 0 0 0 0 0 0 0 0 1 0 0 0\\ 
0 0 0 0 0 0 1 0 0 0 1 0 0 0 1 0 0 0 0 0 0 0 0 0 0 0 0 0 0 0 1 0 0\\ 
0 0 0 0 0 0 0 1 0 0 0 1 0 0 0 1 0 0 0 0 0 0 0 0 0 0 0 0 0 0 1 0 0\\ 
0 0 0 0 1 0 0 0 0 0 0 0 0 0 0 0 1 0 0 0 1 0 0 0 0 0 0 0 0 0 0 1 0\\ 
0 0 0 0 0 1 0 0 0 0 0 0 0 0 0 0 0 1 0 0 0 1 0 0 0 0 0 0 0 0 0 1 0\\ 
0 0 0 0 0 0 1 0 0 0 0 0 0 0 0 0 0 0 1 0 0 0 1 0 0 0 0 0 0 0 0 0 1\\ 
0 0 0 0 0 0 0 1 0 0 0 0 0 0 0 0 0 0 0 1 0 0 0 1 0 0 0 0 0 0 0 0 1\\ 
0 0 0 0 0 0 0 0 0 0 0 0 0 0 0 0 0 0 0 0 0 0 0 0 1 1 1 0 0 1 1 0 0\\ 
0 0 0 0 0 0 0 0 0 0 0 0 0 0 0 0 0 0 0 0 0 0 0 0 1 0 0 1 1 0 0 1 1
    \end{array} \right)
\end{align}
\newpage
\begin{align}
   H_X = \partial_2^G =  \left( \begin{array}{c}
        1 0 0 0 1 0 0 0 1 0 0 0 0 0 0 0 1 0 0 0 0 0 0 0 0 0 0 0 0 0 0 0 0\\ 
0 1 0 0 0 1 0 0 0 1 0 0 0 0 0 0 0 1 0 0 0 0 0 0 0 0 0 0 0 0 0 0 0\\ 
1 0 0 0 0 0 1 0 0 0 1 0 0 0 0 0 0 0 1 0 0 0 0 0 0 0 0 0 0 0 0 0 0\\ 
0 1 0 0 0 0 0 1 0 0 0 1 0 0 0 0 0 0 0 1 0 0 0 0 0 0 0 0 0 0 0 0 0\\ 
0 0 1 0 1 0 0 0 0 0 0 0 1 0 0 0 0 0 0 0 1 0 0 0 0 0 0 0 0 0 0 0 0\\ 
0 0 0 1 0 1 0 0 0 0 0 0 0 1 0 0 0 0 0 0 0 1 0 0 0 0 0 0 0 0 0 0 0\\ 
0 0 1 0 0 0 1 0 0 0 0 0 0 0 1 0 0 0 0 0 0 0 1 0 0 0 0 0 0 0 0 0 0\\ 
0 0 0 1 0 0 0 1 0 0 0 0 0 0 0 1 0 0 0 0 0 0 0 1 0 0 0 0 0 0 0 0 0\\ 
1 1 0 0 0 0 0 0 0 0 0 0 0 0 0 0 0 0 0 0 0 0 0 0 1 1 0 1 0 0 0 0 0\\ 
0 0 1 1 0 0 0 0 0 0 0 0 0 0 0 0 0 0 0 0 0 0 0 0 1 0 1 0 1 0 0 0 0\\ 
0 0 0 0 1 1 0 0 0 0 0 0 0 0 0 0 0 0 0 0 0 0 0 0 1 0 0 0 0 1 0 1 0\\ 
0 0 0 0 0 0 1 1 0 0 0 0 0 0 0 0 0 0 0 0 0 0 0 0 1 0 0 0 0 0 1 0 1\\ 
0 0 0 0 0 0 0 0 1 1 0 0 0 0 0 0 0 0 0 0 0 0 0 0 0 1 0 0 0 1 0 0 0\\ 
0 0 0 0 0 0 0 0 0 0 1 1 0 0 0 0 0 0 0 0 0 0 0 0 0 1 0 0 0 0 1 0 0\\ 
0 0 0 0 0 0 0 0 0 0 0 0 1 1 0 0 0 0 0 0 0 0 0 0 0 0 1 0 0 1 0 0 0\\ 
0 0 0 0 0 0 0 0 0 0 0 0 0 0 1 1 0 0 0 0 0 0 0 0 0 0 1 0 0 0 1 0 0\\ 
0 0 0 0 0 0 0 0 0 0 0 0 0 0 0 0 1 1 0 0 0 0 0 0 0 0 0 1 0 0 0 1 0\\ 
0 0 0 0 0 0 0 0 0 0 0 0 0 0 0 0 0 0 1 1 0 0 0 0 0 0 0 1 0 0 0 0 1\\ 
0 0 0 0 0 0 0 0 0 0 0 0 0 0 0 0 0 0 0 0 1 1 0 0 0 0 0 0 1 0 0 1 0\\ 
0 0 0 0 0 0 0 0 0 0 0 0 0 0 0 0 0 0 0 0 0 0 1 1 0 0 0 0 1 0 0 0 1
    \end{array} \right)
\end{align}

As we now have a 5-term chain complex, we get metachecks $M_Z, M_X$ for both the $Z$- and $X$-type syndromes. However, both classical codes formed from these $M_Z$ and $M_X$ have distance $d=2$, and so the only thing we can do with them is detect a single syndrome error. Using $L > 2$ would give us the ability to actually correct syndrome errors before decoding normally.
\begin{align}
   M_Z^T = \partial_4^G =  \left( \begin{array}{c}
        1 0 1 0 1 0 0 0 0 0 1 0 1 0 0 0 0 0 1 0\\
0 1 0 1 0 1 0 0 0 0 0 1 0 1 0 0 0 0 1 0\\
1 0 0 0 0 0 1 0 1 0 0 0 0 0 1 0 1 0 0 1\\
0 1 0 0 0 0 0 1 0 1 0 0 0 0 0 1 0 1 0 1
    \end{array} \right)
\end{align}

\begin{align}
   M_X = \partial_1^G =  \left( \begin{array}{c}
        1 1 0 0 0 0 0 0 1 0 1 0 1 0 0 0 1 0 0 0\\
0 0 1 1 0 0 0 0 1 0 0 1 0 1 0 0 0 1 0 0\\
0 0 0 0 1 1 0 0 0 1 1 0 0 0 1 0 0 0 1 0\\
0 0 0 0 0 0 1 1 0 1 0 1 0 0 0 1 0 0 0 1
    \end{array} \right)
\end{align}

\end{document}